# Reversible 300K Ferromagnetic Ordering in a Diluted Magnetic Semiconductor**

*Dana A. Schwartz and Daniel R. Gamelin**

**Abstract.** The discovery of reversible 300 K ferromagnetic ordering in a diluted magnetic semiconductor is reported. Switching of room-temperature ferromagnetism between "on" and "off" states is achieved in $Co^{2+}$:ZnO by lattice incorporation and removal of the native n-type defect, interstitial Zn. Spectroscopic and magnetic data implicate a double-exchange mechanism for ferromagnetism. These results demonstrate for the first time reversible room-temperature ferromagnetic ordering in a diluted magnetic semiconductor, and present new opportunities for integrating magnetism and conductivity in semiconductor sensor or spin-based electronics devices.

* Prof. D. R. Gamelin, Dr. D. A. Schwartz
  Department of Chemistry
  University of Washington
  Seattle, WA 98195-1700 (USA)
  E-mail: Gamelin@chem.washington.edu

** This work was funded by the NSF (DMR-0239325 and ECS-0224138), the Semiconductor Research Corporation (2002-RJ-1051G), and the UW/PNNL Joint Institutes for Nanoscience (Graduate Research Award to D.A.S.). D.R.G. is a Cottrell Scholar of the Research Corporation. We thank Profs. S. A. Chambers and K. M. Krishnan for donation of materials.

Diluted magnetic semiconductors (DMSs) are the active components of many proposed devices designed to process information by manipulating the spins of electrical currents (spintronics).[1, 2] Since the recent discovery of room-temperature ferromagnetism in cobalt-doped $TiO_2$,[3] several reports of high-$T_C$ ferromagnetic DMSs have appeared, including DMSs of ZnO,[4-7] $SnO_2$,[8] and GaN.[9, 10] These new materials offer hope for the development of practical semiconductor spintronics technologies. Little is understood about the origins of ferromagnetism in these high-$T_C$ DMSs,[9-11] however, partly because of the important roles played by defects and partly because most show n-type conductivity, in contrast with the widely investigated lower-$T_C$ p-type DMSs.[12-14] Reproducibilities have generally been low, and in many cases even the claim of intrinsic ferromagnetism remains contentious. Here we report the discovery of reversible 300 K ferromagnetic ordering in $Co^{2+}$:ZnO, switched between "on" and "off" states with quantitative reproducibility by introducing and removing interstitial Zn ($Zn_i$), a native n-type defect of ZnO. These results represent the first reversible room-temperature ferromagnetic ordering demonstrated for any DMS, a discovery with important implications for spintronics technologies. The highly reproducible spectroscopic and magnetic changes that accompany switching implicate a double-exchange mechanism for magnetic ordering that involves electron delocalization within a sub-conduction-band cobalt impurity level.

To test the hypothesis that n-type defects promote carrier-mediated ferromagnetism in ZnO DMSs,[15] we have explored the influence of $Zn_i$ on the physical properties of $Co^{2+}$:ZnO thin films. Experiment and theory suggest that $Zn_i$ is a shallow (~0.030 eV) electron donor in ZnO and is predominantly responsible for its native n-type character.[16, 17] $Zn_i$ is readily formed at low concentrations as a growth defect, and its concentration can be increased up to ~$10^{20}$/cm$^3$ by Zn metal vapor diffusion.[18, 19] Fig. 1 shows electronic absorption spectra of two representative 3.61% ($1.53 \times 10^{21}/cm^3$) $Co^{2+}$:ZnO thin films of 0.50 and 0.05 μm thicknesses, prepared by spin-coating colloidal DMS nanocrystals onto fused silica substrates. These absorption spectra are characteristic of ZnO, which has a 300 K band gap of ~3.4 eV. The thinner film (Fig. 1a) allows the first excitonic maximum (~3.60 eV) to be observed. The spectrum of the thicker film (Fig. 1b) is essentially identical but is distorted by stray light above ~3.5 eV. The spectrum of Fig. 1b plotted on an expanded scale (inset) shows a structured absorption band centered at ~1.98 eV, identified as the spin-orbit split $^4A_2 \rightarrow {}^4T_1(P)$ ligand-field transition of $Co^{2+}$ ions substitutionally doped into wurtzite ZnO.[6, 20] Upon exposure of the film to Zn vapor for 5 hours at 500 °C, the band gap absorbance decreased by ~10% and a new absorption feature of comparable intensity appeared in the infrared. This strong infrared absorption has been reported previously for Zn-treated ZnO[19] and is characteristic of free carriers in heavily n-doped semiconductors. Very similar spectral changes were also reported recently for colloidal ZnO nanocrystals when reduced by strong chemical reductants.[21] The data in Fig. 1 are thus consistent with the introduction of n-type carriers into ZnO DMS films by $Zn_i$ doping. Concomitant with these spectral changes is a 21% decrease in the $Co^{2+}$ ligand-field absorbance (Fig. 1b, inset) and appearance of new sub-bandgap absorption intensity that tails into the visible.

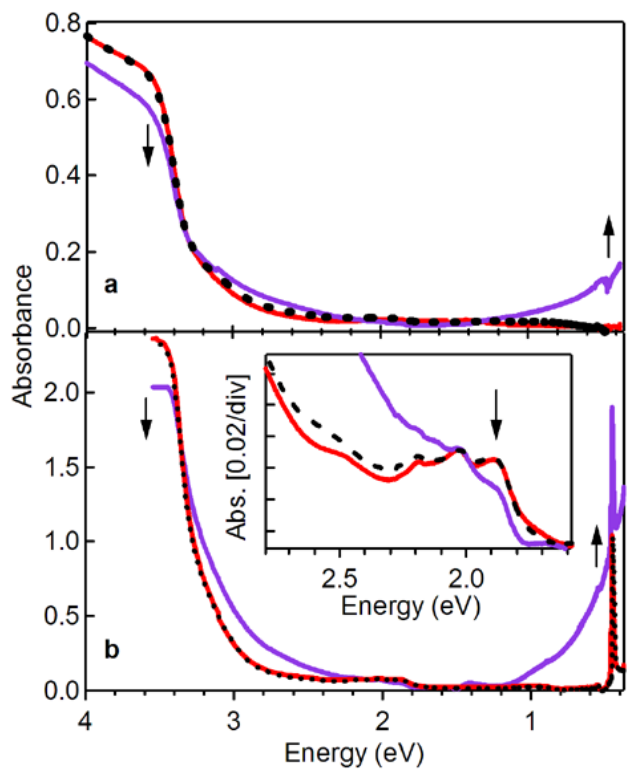

**Figure 1.** Electronic absorption spectra of 3.61% $Co^{2+}$:ZnO thin films prepared by spin coating DMS nanocrystals onto fused silica substrates: **(a)** 0.05 μm and **(b)** 0.50 μm thickness. Data for both films: As-prepared (red), following 5 hrs exposure to Zn vapor at 500 °C (blue), and subsequently exposed to air for ≥2 min at 500 °C (black dashed). Inset: $Co^{2+}$ ligand-field region of the spectra from (b) shown on an expanded scale. The arrows indicate changes upon exposure to Zn vapor.

Analysis of the Burstein-Moss shift[22-24] in ZnO band gap energies from Fig. 1 shows the free carrier concentration was increased to ca. $2 \times 10^{19}/cm^3$ upon exposure to Zn vapor. This increase was confirmed by four-point resistivity measurements. The resistivity of the film from Fig. 1b was initially >$10^7$ Ω cm. Upon exposure to Zn vapor the resistivity decreased to 0.3 Ω cm. Parallel studies on $Co^{2+}$:ZnO thin films grown epitaxially by MOCVD verify the trend ($10^{-3}$ Ω cm before, $10^{-5}$ Ω cm after, Figs. S5-S7). This decrease in resistivity is attributed to incorporation of the shallow donor $Zn_i$ into the DMS lattice.

Fig. 2 plots magnetic susceptibility data collected on the same film from Fig. 1b. Before exposure to zinc vapor, the film was paramagnetic with no ferromagnetic or superparamagnetic signal detected at any temperature between 10 and 350 K. The 300 K data are shown in Fig. 2. Upon exposure to Zn vapor, a clear ferromagnetic hysteresis was observed that persisted up to our experimental limit of 350 K (Figs. 2 and S1). Control experiments performed by heating the films without Zn metal or with Sn metal in place of Zn metal under identical conditions showed no significant change in magnetism, conductivity, or electronic absorption, confirming attribution of the physical changes to Zn vapor. The 300 K ferromagnetic saturation moment ($M_S$) of the film in Fig. 2 was 0.046 $\mu_B$/cobalt and could be increased to 0.079 $\mu_B$/cobalt by extending the Zn vapor exposure time to 9 hours, beyond which it decreased (Fig. 2 inset). Although relatively small in absolute magnitude, the induced ferromagnetism is commensurate with the small concentration of $Zn_i$ introduced by vapor diffusion. Essentially identical spectroscopic, magnetic, and conductivity changes were obtained for $Co^{2+}$:ZnO thin films grown epitaxially by MOCVD (Figs. S5-S7), demonstrating that the effect is not related to film preparation or surface area. Once prepared, the physical properties of the Zn-treated films showed no decay after storage in air at room temperature for several months. We emphasize that the activation of 300 K ferromagnetism shown in Fig. 2 arises from well-controlled introduction of a known lattice defect. Consequently, the activation is quantitatively reproducible and has been successful on 100% of our attempts (15 films).

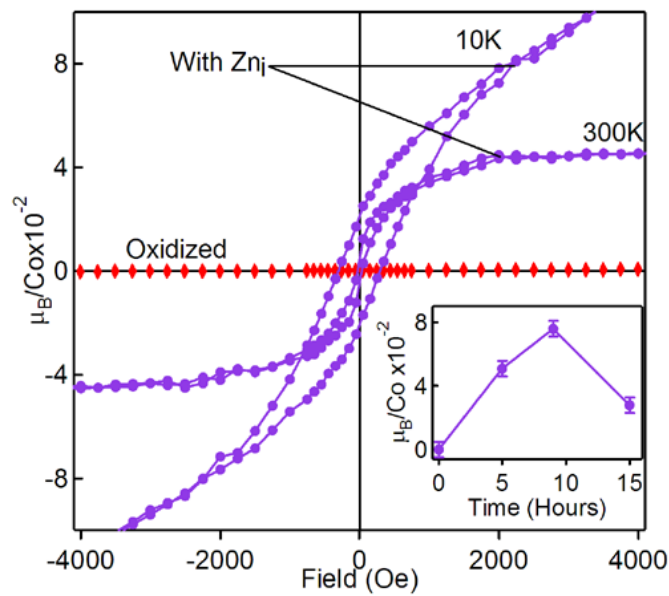

**Figure 2.** Magnetic susceptibility of the 3.61% $Co^{2+}$:ZnO thin film from Fig. 1b. Red: 300 K susceptibility of the as-prepared (oxidized) film. Blue: 10 and 300 K susceptibility of the same film following exposure to Zn vapor. Inset: Dependence of the 300 K saturation moment measured at 3000 Oe on the duration of Zn vapor exposure. All data have been corrected for substrate diamagnetism.

The correlation between $Zn_i$ incorporation, conductivity, and ferromagnetism is consistent with a carrier-mediated mechanism for magnetic ordering.[13, 15] Since $Zn_i$ is an electron donor, these data show that electrons are the effective carriers, in contrast with the hole-mediated ferromagnetism demonstrated in some DMSs at low temperatures.[12] The bleaching of the $Co^{2+}$ ligand-field absorption (Fig. 1b, inset) further show that electrons introduced into $Co^{2+}$:ZnO interact strongly with the $Co^{2+}$ dopants. Electron trapping would be consistent with reports that the carrier density of ZnO decreases with $Co^{2+}$ doping.[4] The potential of the $Co^+$ ground state relative to the conduction band (CB) minimum may be estimated from the energy of the valence band (VB) → $Co^{2+}$ electronic transition, assigned previously using absorption and magnetic circular dichroism spectroscopies.[6] As shown in Fig. 3a, this transition overlaps the band gap transition of ZnO (VB → CB) and extends to slightly lower energies. Neglecting possible oxo polarization effects in the immediate vicinity of the cobalt ion, the energy difference between these two transitions reflects the potential difference between the CB and the $Co^+$ ground state in $Co^+$:ZnO. From these data, an upper limit of 0.3 eV is estimated for this difference, and electron transfer from $Zn_i$ to $Co^{2+}$ should be energetically favorable (Fig. 3b). The cooperative magnetic ordering that results clearly indicates that the extra electrons are not localized at individual $Co^+$ ions, however, and this conclusion is supported by observation of the infrared absorption characteristic of free carriers in the DMS (Fig. 1). We propose that the extra electrons are delocalized among several (n) cobalt ions to form a shallow impurity band. The driving force for delocalization of the extra electron (Fig. 3b) is spin dependent, giving rise to the maximum stabilization when the magnetic moments of all n cobalt ions are aligned in a net ferromagnetic configuration. This delocalization-induced ferromagnetism, referred to as double exchange[25, 26] and illustrated schematically in Fig. 3c for the limiting case of n = 2, was previously reported for $Mn^{3+/4+}$ perovskites[25-27] and certain transition-metal clusters,[28] and proposed for $Co^{2+}$:ZnO on the basis of density functional calculations.[15] Itinerant electrons are therefore directly responsible for the 300 K ferromagnetism observed in this DMS. We speculate that the double-exchange resonance integral governing delocalization is enhanced by the energetic proximity of the $Co^+$ ground state to the ZnO CB, which favors hybridization of the $Co^+$ and CB electronic wavefunctions as described by perturbation theory. This hybridization is in addition to the substantial p-d hybridization with the valence band[6, 13] and contributes an effective orbital pathway for electron delocalization among cobalt ions.

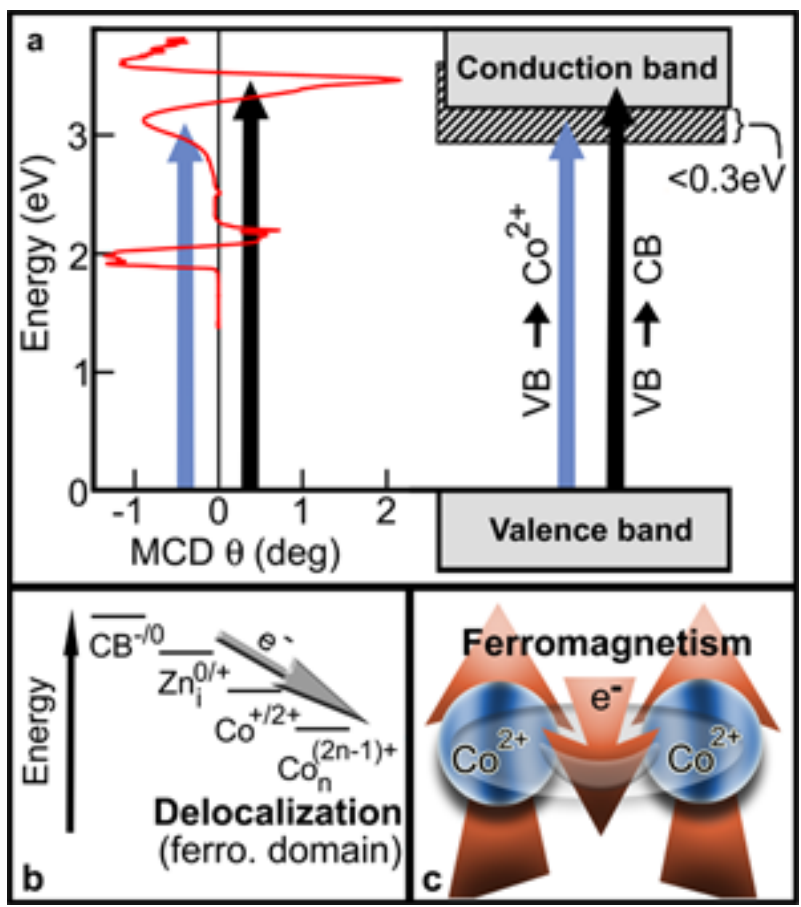

**Figure 3.** **(a)** Relationship between VB → $Co^{2+}$ charge-transfer transition energy, VB → CB transition energy, and $Co^{+/2+}$ trap energy (< 0.3eV) in $Co^{2+}$:ZnO. MCD spectrum adapted from ref. [6]. **(b)** Schematic illustration of the thermodynamic driving force for electron transfer from $Zn_i$ to $Co^{2+}$, and for subsequent electron delocalization among n $Co^{2+}$ ions giving rise to ferromagnetism by double exchange. **(c)** Schematic illustration of double-exchange ferromagnetism involving d-electron itinerancy in a $Co_n^{(2n-1)+}$ magnetic unit for the limiting case of n = 2.

Perhaps the most remarkable aspect of this room-temperature magnetic ordering is its reversibility. Heating the Zn-treated thin film from Fig. 1b in air rapidly quenched its ferromagnetism (Fig. 4a) and returned it to its original paramagnetic state. The oxidation was complete after only ca. 2 minutes in air at 500 °C (Fig. S2). Furthermore, re-exposure of the same film to Zn vapor returned it quantitatively to the ferromagnetic state (Fig. 4a). Fig. 4b quantifies the changes in ferromagnetism ($M_S$) at 300 K, the $Co^{2+}$ $^4A_2 \rightarrow {}^4T_1(P)$ ligand-field absorption intensity, the conductivity ($\Omega^{-1}$ $cm^{-1}$), and the 10 K $^4A_2$ paramagnetic susceptibility for the film used in Figs. 1b and 2. Upon exposure to Zn vapor, the ligand-field absorption intensity and paramagnetism both decrease by 21% and the conductivity and ferromagnetism appear. Upon oxidation, the $^4A_2 \rightarrow {}^4T_1(P)$ ligand-field absorption and $^4A_2$ paramagnetism return quantitatively to their pre-reduction values and the film is no longer ferromagnetic or conductive. A second Zn treatment causes all of the measured parameters to return quantitatively to the same values they had after the first reduction. The chemical perturbation that leads to room-temperature ferromagnetism in this $Co^{2+}$:ZnO thin film

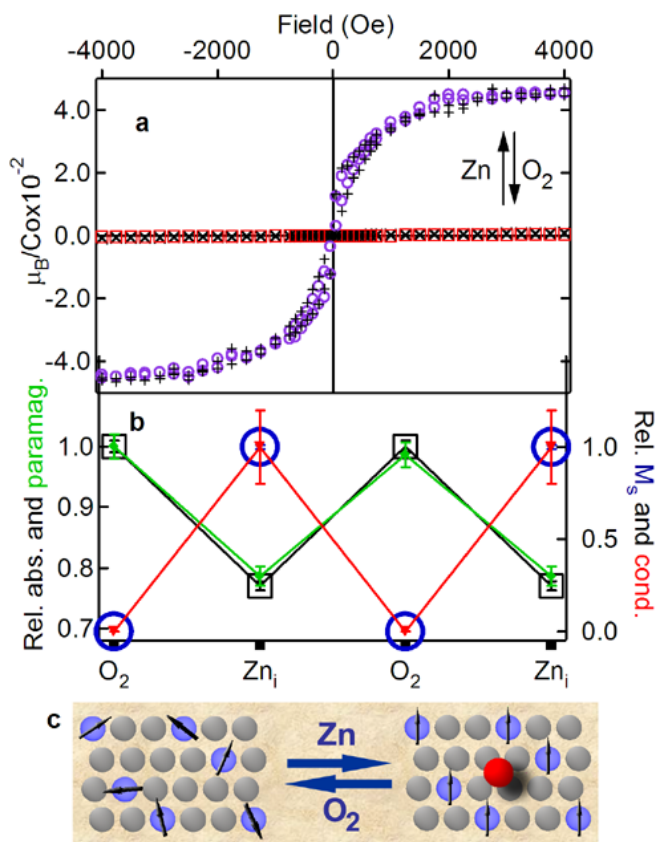

**Figure 4.** **(a)** Interconversion of 300 K ferromagnetism between "on" and "off" states for the 3.61% $Co^{2+}$:ZnO thin film from Figs. 1b and 2 using alternating $Zn_i$ incorporation and oxidation with $O_2$ (Sequence: × ($O_2$), ○ ($Zn_i$), □ ($O_2$), + ($Zn_i$)). **(b)** Correlated switching of the 300 K $^4A_2 \rightarrow {}^4T_1(P)$ ligand-field absorption intensity (black), 10 K $Co^{2+}$ paramagnetism (green), 300 K ferromagnetic saturation moment, $M_S$ (blue), and 300 K conductivity (red) for the same film. **(c)** Schematic illustration of the effect of $Zn_i$ on magnetic ordering.

41111

is therefore fully and quantitatively reversible. Because of this quantitative reversibility, particularly as monitored by ligand-field absorption spectroscopy (Fig. 1, inset), this switching process is highly unlikely to arise from phase segregation, since that would require quantitative re-incorporation of segregated cobalt back into the ZnO lattice in under 2 minutes at 500 °C (Fig. S2). The very similar results obtained for $Co^{2+}$:ZnO films grown epitaxially by MOCVD (Figs. S5-S7) strongly support this conclusion by demonstrating that the results are independent of the method of $Co^{2+}$:ZnO preparation or the surface area of the film. Furthermore, the thin films showed no change in magnetism or absorption when annealed at 500 °C, $10^{-3}$ Torr for 15 hours in the absence of Zn, indicating they are stable against cobalt segregation under these conditions. Finally, control experiments on Co metal nanocrystals performed under identical conditions showed largely irreversible oxidation. We conclude that the switching arises from incorporation of $Zn_i$ into the lattice of paramagnetic $Co^{2+}$:ZnO, which activates ferromagnetism by donating electrons, and oxidation with $O_2$, which deactivates ferromagnetism by removing those electrons. This switching process is summarized schematically in Fig. 4c. Importantly, although Zn and $O_2$ were used here, because the switching ultimately relates to introduction and removal of electrons its extension to a variety of other redox chemical, photochemical, or electrical conditions may be anticipated, suggesting rich opportunities for integrating magnetic switching and conductivity in semiconductor sensor or spin-electronics devices.

In summary, we have demonstrated the first reversible 300 K ferromagnetic ordering in a DMS, achieved in $Co^{2+}$:ZnO by lattice incorporation and removal of the native n-type defect, $Zn_i$. Spectroscopic and magnetic data implicate a double-exchange mechanism for ferromagnetism that involves electron delocalization among substitutionally doped cobalt ions. The discovery of reversible 300 K ferromagnetic ordering in a semiconductor presents new opportunities for spin-based device applications that may considerably impact future information processing technologies.

*Experimental*

Thin films (0.50 and 0.05 μm thickness) of $Co^{2+}$:ZnO were prepared by spin coating colloidal dodecylamine-capped 7 nm diameter 3.61 ± 0.06% $Co^{2+}$:ZnO nanocrystals onto fused silica or water-free quartz substrates (20 and 2 coats, respectively) and heating at 500 °C in air for 2 minutes to remove solvent and ligands. The nanocrystals were prepared, purified, and characterized in detail as described previously.[6] Cobalt concentrations were determined by inductively coupled plasma atomic emission spectroscopy. The $Co^{2+}$:ZnO films were exposed to zinc vapor by heating with metallic Zn to 500 °C in a quartz tube at $5 \times 10^{-3}$ Torr. Electronic absorption, magnetic susceptibility, and X-ray diffraction data were collected as described previously.[6] Resistivities were measured using an in-line four-point probe. The epitaxial thin film of 9% $Co^{2+}$:ZnO prepared by MOCVD was graciously provided by S. A. Chambers, and the Co metal nanocrystals for control experiments were provided by K. M. Krishnan.